\documentclass[a4paper,fleqn,usenatbib]{mnras}

\usepackage{newtxtext,newtxmath}

\usepackage[T1]{fontenc}
\usepackage{ae,aecompl}

\usepackage{graphicx}	
\usepackage{amsmath}	
\usepackage{amssymb}	

\graphicspath{{images/}}

\usepackage{xcolor}
\newcommand{\revision}[1]{#1}

\title[The PSF as convolutive orders of the phase PSD]{Turbulent and adaptive optics corrected point-spread functions as convolutive orders of the phase power spectral density}
\author[R. JL. F\'etick et al.]{
Romain JL. F\'etick,$^{1,2}$\thanks{E-mail: romain.fetick@lam.fr}
Benoit Neichel,$^{1}$
Laurent \revision{M} Mugnier,$^{2}$
\newauthor
Aur\'elie Montmerle-Bonnefois,$^{2}$
and Thierry Fusco$^{1,2}$
\\
$^{1}$Aix Marseille Univ, CNRS, CNES, LAM, Marseille, France\\
$^{2}$ONERA, The French Aerospace Lab BP72, 29 avenue de la Division Leclerc, 92322 Ch\^atillon Cedex, France\\
}

\date{Accepted 2018 September 13. Received 2018 September 13; in original form 2018 July 20}

\pubyear{2018}

\begin{document}
\label{firstpage}
\pagerange{\pageref{firstpage}--\pageref{lastpage}}
\maketitle

\begin{abstract}
Ground-based astronomy is severely limited by the atmospheric turbulence, resulting in a large Point-Spread-Function (PSF) and poor imaging resolution. Even imaging with Adaptive Optics (AO) cannot completely correct the aberrated wavefront, and a residual turbulence still corrupts the observation. Thus the consequences of the turbulence on the PSF is of first interest when building any ground-based telescope. The Power Spectral Density (PSD) of a spatially stationary turbulent phase carries all the information needed for describing the long-exposure PSF. We then develop an analytical description of the long-exposure PSF as a series expansion of the aberrated phase PSD. Our description of the PSF given the PSD of the phase is a simple theoretical way to describe the impact of turbulence on the PSF. We also show accordance with previous papers when restricting our model to its first expansion order. Finally we derive applications of our formula to some particular cases, such as Kolmogorov or von-K\'arm\'an models, or the AO correction impact on the PSF.
\end{abstract}




\section{Introduction}

The Point Spread Function is a critical information for any optical system since it is directly related to the imaging resolution. Diffraction limited optics with a circular aperture results in the famous Airy pattern. However, in practice, the wavefront of the observed scene suffers from phase aberrations during its propagation through an inhomogeneous medium, reducing imaging performance. In case of ground-based telescopes, the atmospheric turbulence distorts the wavefront and severely limits the resolution of telescopes \citep{roddier1981v}. Adaptive Optics (AO) is a technique using deformable mirrors to reduce phase aberrations by flattening the wavefront back \citep{roddier1999adaptive}. Nevertheless the AO correction is not perfect and some phase aberrations remain, especially the high spatial frequencies which are not seen by the wavefront sensor and thus not corrected.\\

On \revision{the} one hand, \citet{roddier1981v} developed a relation to find the long-exposure PSF from the phase Power Spectral Density (PSD). The formula is exact but the link PSD/PSF is not direct since it requires a double Fourier transform. On the other hand \citet{racine1999speckle,bloemhof2001behavior,jolissaint2002fast} derived a first order approximation of the \citeauthor{roddier1981v} method. The advantage is to explain the direct link between PSD and PSF, by canceling the double Fourier transformation, however the drawback is an inaccuracy due to the first order approximation.\\

The two methods above \revision{illustrate} the very close relation between the phase PSD and the PSF, each one with its advantages and drawbacks. In this paper we use the \citeauthor{roddier1981v} method to derive an exact expression of the PSF, with a direct analytical link PSD/PSF without requiring the multiple Fourier transforms. Indeed, our method relies on a convolutional formalism, where the PSF is described as an infinite sum of convolutive orders of the phase PSD. Finally, when restricting our expression to its first order, we find the approximations of \citeauthor{racine1999speckle,bloemhof2001behavior,jolissaint2002fast}.\\

Straightforward applications involve practical cases where the turbulent, or partially corrected AO, phase is not known for each time-step, but its PSD follows well-known statistics (e.g. Kolmogorov, von-K\'arm\'an law, AO residuals PSD). As a consequence our model can be easily fed with available information on the turbulence. Our direct expression between PSD and PSF also allows for simple estimation of atmospheric parameters directly from the PSF.\\

Section \ref{sec:method} develops our analytical method that writes the PSF as orders of the residual aberrated phase PSD. This method is general and can be used for any spatially stationary phase. Section \ref{sec:applications} shows some applications related to Kolmogorov or von-K\'arm\'an statistics. We give an explicit dependence of the PSF shape with the Fried parameter $r_0$ \citep{fried1966optical}. We also derive a simple description of the halo for long-exposure PSF using AO correction. Section \ref{sec:conclusion} concludes our work. We discuss about future applications of our model.


\section{Method}
\label{sec:method}

\subsection{Factorization of the OTF}

Let's call $P$ the entrance pupil transmission function \revision{(that may include static aberrations)}, $h$ the long-exposure PSF and $\psi$ the complex phasor \citep{goodman1968fourier} \revision{that is the complex scalar field when observing a point source. The scintillation is neglected so the complex phasor has unit amplitude.} The long-exposure Optical Transfer Function (OTF) is given by the autocorrelation of the phasor on the pupil (\citeauthor{goodman1968fourier})
\begin{equation}
\tilde{\mathrm{h}}(\vec{\rho}/\lambda) = \frac{1}{S}\iint P(\vec{r})P(\vec{r}+\vec{\rho})\langle\psi (\vec{r},t)\psi (\vec{r}+\vec{\rho},t)\rangle_t \mathrm{d}\vec{r}  
\end{equation}
where $S$ ensures normalization on the surface of the pupil,  $\langle\cdot\rangle_t$ denotes the time average and $\tilde{h}$ stands for the Fourier transform of the PSF $h$. We then write the phasor $\psi$ as the complex exponential of the aberrated phase $\phi$
\begin{equation}
\psi(\vec{r},t) = e^{i\phi(\vec{r},t)}
\end{equation}

\citet{roddier1981v} has shown that assuming Gaussian statistics of the phase, the long-exposure OTF writes
\begin{equation}
\tilde{\mathrm{h}}(\vec{\rho}/\lambda) = \frac{1}{S}\iint P(\vec{r})P(\vec{r}+\vec{\rho}) e^{-\frac{1}{2}D_\phi(\vec{r},\vec{\rho})} \mathrm{d}\vec{r} 
\end{equation}
where $D_\phi(\vec{r},\vec{\rho})$ is the structure function of the residual phase, defined as
\begin{equation}
D_\phi(\vec{r},\vec{\rho}) = \langle |\phi(\vec{r},t)-\phi(\vec{r}+\vec{\rho},t)|^2\rangle_t
\end{equation}

Assuming spatially stationary residual phase, the dependency in the position $\vec{r}$ vanishes and only remains the separation $\vec{\rho}$. This stationarity hypothesis is verified for a purely tubulent phase, and is a reasonable approximation in case of partial AO correction \citep{conan1994etude}. The approximation is better for larger aperture telescopes, since the non-stationarity of the residual phase essentially affects the edges of the pupil. The structure function then simplifies
\begin{equation}
D_\phi(\vec{r},\vec{\rho}) = D_\phi(\vec{\rho}) = \langle\langle |\phi(\vec{r},t)-\phi(\vec{r}+\vec{\rho},t)|^2\rangle_t\rangle_r
\end{equation}

The OTF can then be separated into two OTFs, the telescope one and the atmospheric one
\begin{equation}
\label{eq:OTF_tel_atm}
\tilde{\mathrm{h}}(\vec{\rho}/\lambda) = \tilde{\mathrm{h}}_T(\vec{\rho}/\lambda) \cdot \tilde{\mathrm{h}}_A(\vec{\rho}/\lambda)
\end{equation}
with the telescope OTF
\begin{equation}
\tilde{\mathrm{h}}_T(\vec{\rho}/\lambda) = \frac{1}{S}\iint P(\vec{r})P(\vec{r}+\vec{\rho}) \mathrm{d}\vec{r}  
\end{equation}
and the atmospheric OTF
\begin{equation}
\label{eq:OTFatmo}
\tilde{\mathrm{h}}_A(\vec{\rho}/\lambda) = e^{-\frac{1}{2}D_\phi(\vec{\rho})}
\end{equation}
Even though this expression may give the PSF by a numerical Fourier transform, it doesn't \revision{exhibit an} explicit dependence in the PSD \revision{of the phase}. It is then necessary to modify the equation to find the direct link we are looking for.

\subsection{Expression of the PSF}

Assuming the phase variance is finite, let's define the autocorrelation of the phase
\begin{equation}
B_\phi(\vec{\rho}) = \langle\langle \phi(\vec{r},t)\phi(\vec{r}+\vec{\rho},t) \rangle_t\rangle_r
\end{equation}
Directly follows the equality
\begin{equation}
D_\phi (\vec{\rho}) = 2 [B_\phi (\vec{0}) - B_\phi (\vec{\rho})]
\end{equation}
Using this identity and the series expansion of the exponential, one gets the atmospheric OTF
\begin{equation}
\tilde{\mathrm{h}}_A(\vec{\rho}/\lambda) = e^{-B_\phi (\vec{0})} e^{B_\phi (\vec{\rho})} = e^{-B_\phi (\vec{0})}\sum_{n=0}^{+\infty}\frac{B_\phi (\vec{\rho})^n}{n!}
\end{equation}
We denote $\mathcal{F}\{f\}(\vec{u})$ the Fourier transform operator applied on the function $f$, evaluated in $\vec{u}$. Using the linearity of the Fourier operator on the infinite sum, the PSF writes
\begin{equation}
\mathrm{h}_A(-\vec{u}) = e^{-B_\phi (\vec{0})}\sum_{n=0}^{+\infty}\frac{\mathcal{F}\{B_\phi (\vec{\rho})^n\}(\vec{u})}{n!}
\end{equation}
where $\vec{\rho}/\lambda$ and $\vec{u}$ are conjugated variable by the Fourier transform. Since the Fourier transform links multiplication and convolution, we define the convolutive orders of a function $W$ as
\begin{equation}
\label{eq:def_conv}
\left\{ \star^n~ W \right\} (u) = 
\left\lbrace
\begin{array}{cl}
\delta (u) & \text{ if }n=0\\
W (u) & \text{ if }n=1\\
(W\star W\star ... \star W)(\vec{u}) & \text{ if }n\geq 2\\
\end{array}
\right.
\end{equation}
where $\delta(\vec{u})$ denotes the Dirac distribution. Then the PSF writes
\begin{equation}
\mathrm{h}_A(-\vec{u}) = e^{-B_\phi (\vec{0})}\sum_{n=0}^{+\infty}\frac{\star^n~\mathcal{F}\{B_\phi (\vec{\rho})\}(\vec{u})}{n!}
\end{equation}
By the Wiener-Khintchine theorem, the Fourier transform of the autocorrelation is equal to the PSD of the phase. We write $W_\phi$ the PSD of the phase, and taking care of coordinates \revision{dilation} for the Fourier transform
\begin{equation}
\label{eq:PSF_as_PSD}
\mathrm{h}_A(-\vec{u}) = e^{-B_\phi (\vec{0})}\sum_{n=0}^{+\infty}\frac{ \left\{ \star^n~ W_\phi/\lambda^2 \right\} (\vec{u}/\lambda)}{n!}
\end{equation}
This equation states that the atmospheric long-exposure PSF can be written as an infinite sum of convolutive orders \revision{of the phase} PSD, under assumption of spatially stationary phase. In other words, the long-exposure PSF can be retrieved when the turbulent phase for each time-step is unknown but its PSD is well-known. This is the case for long-exposure observation through the turbulent atmosphere, without or with adaptive optics. Then, using Eq. (\ref{eq:OTF_tel_atm}) to retrieve the total PSF, atmospheric and diffraction, directly writes
\begin{equation}
\label{eq:PSF_tel_atm}
h(\vec{u}) = h_T(\vec{u}) \star h_A(\vec{u})
\end{equation}

The convolutive formalism of Eq. \ref{eq:PSF_as_PSD} can be found in the literature \citep{sivaramakrishnan2002speckle,perrin2003structure}, but for short-exposure PSF only. These authors described the short-exposure PSF as a double infinite sum on the instantaneous phase. As we show in this paper, based on Roddier's method, the long-exposure PSF allows simplifications to a unique infinite sum on the phase PSD.


\section{Applications}
\label{sec:applications}

\subsection{Factorization of the $r_0$ dependency}
\label{sec:sub:r0}

\begin{figure}
   \centering
   \includegraphics[width=7cm]{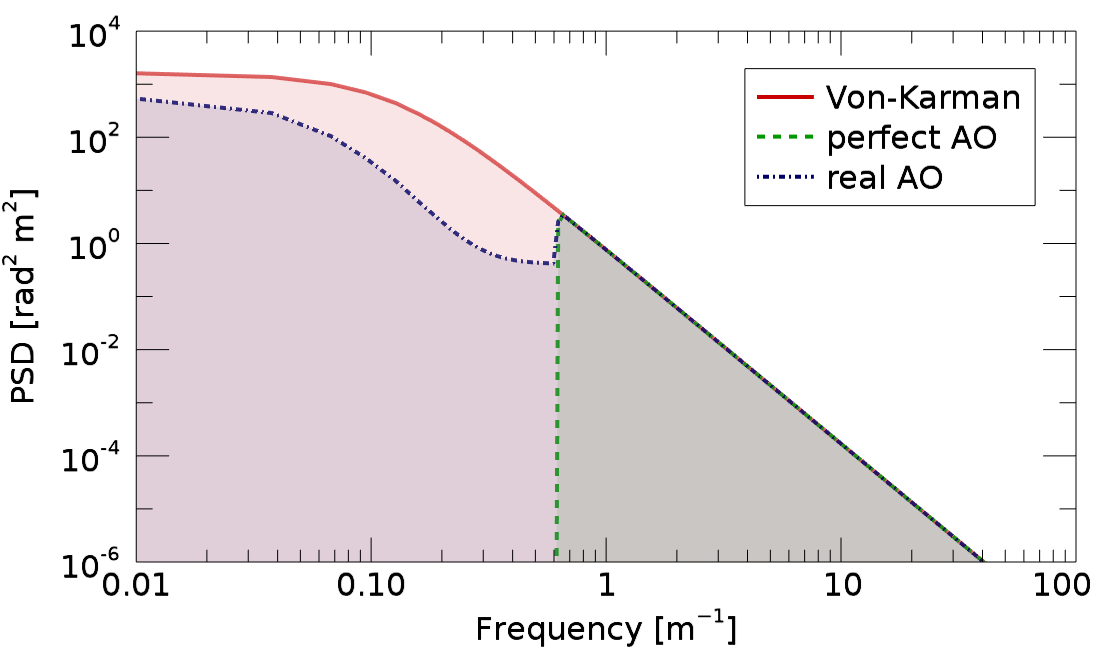}
   \caption{Three models of \revision{phase} PSD considered in the following applications: von-K\'arm\'an (red), perfect AO as truncated Kolmogorov (green) and more realistic AO correction (blue).}
\label{fig:PSD_models}
\end{figure}

For usual turbulence spectrum such as Kolmogorov or von-K\'arm\'an, there is a direct dependency of the PSF in the Fried parameter $r_0$. Let's consider the von-K\'arm\'an spectrum
\begin{equation}
\label{eq:VK}
W_\phi(f) = 0.023 r_0^{-5/3}\left( \frac{1}{L_0^2} + f^2 \right)^{-11/6}
\end{equation}
where $L_0$ is the external scale of the turbulence. The von-K\'arm\'an spectrum for $L_0=8$ m is plotted \revision{on} Fig. \ref{fig:PSD_models}. Letting $L_0\to\infty$ one would find the Kolmogorov spectrum.\\

Let's now extract the $r_0$ dependency from the PSD. We define $Y_\phi$ the PSD of the phase with unit $r_0$. Then, by definition of $Y_\phi$
\begin{equation}
\frac{W_\phi(f)}{\lambda^2} = r_0^{-5/3} Y_\phi(f)
\end{equation}
The convolutive orders of $Y_\phi$ can be numerically computed only once for a given $L_0$. The PSF for different values of $r_0$ is then a weighted sum of these components. Plugging $Y_\phi$ into Eq. (\ref{eq:PSF_as_PSD}) gives the PSF
\begin{equation}
\mathrm{h}_A(-\vec{u}) = e^{-B_\phi (\vec{0})}\sum_{n=0}^{+\infty}\frac{r_0^{-5n/3}}{n!} \left\{ \star^n~ Y_\phi \right\} (\vec{u}/\lambda)
\end{equation}

The impact of the strength of the turbulence on the PSF is then more explicit: for high $r_0$ (weak turbulence) the high order convolutive terms have little importance. Consequently the shape of the PSF is dominated by the low convolutive orders. For low $r_0$ (strong turbulence) the higher convolutive orders get more importance for describing the PSF shape.\\

For a practical application, the infinite sum is truncated up to a number $N_\mathrm{max}$ of orders. One would like to get an idea on this number of terms to consider for having an accurate PSF model. The number $N_\mathrm{max}$ depends on $r_0$ but also on the evolution of the $Y_\phi$ convolutive orders with $n$. Let's call $E$ the integral of $Y_\phi$, or equivalently the phase variance for a unit $r_0$. We find a \revision{semi-}empirical approximation \revision{(see explanations in appendix \ref{appendix:approx}) for values of $\vec{u}$} in the halo
\begin{equation}
\left\{ \star^n~ Y_\phi \right\} (\vec{u}/\lambda) \approx E^{n-1} Y_\phi (\vec{u}/\lambda)
\end{equation}

Dropping the term $n=0$ (particular behavior of the Dirac in $\vec{u}=\vec{0}$), we get a rough estimate of the PSF behavior towards the orders
\begin{equation}
\forall \vec{u}\in \{ \text{halo} \} ~ , ~~ \mathrm{h}_A(-\vec{u}) \approx  Y_\phi(\vec{u}/\lambda) e^{-B_\phi (\vec{0})} \sum_{n=1}^{+\infty}\frac{E^{n-1} r_0^{-5n/3}}{n!}
\end{equation}

\begin{figure}
   \centering
   \includegraphics[width=7cm]{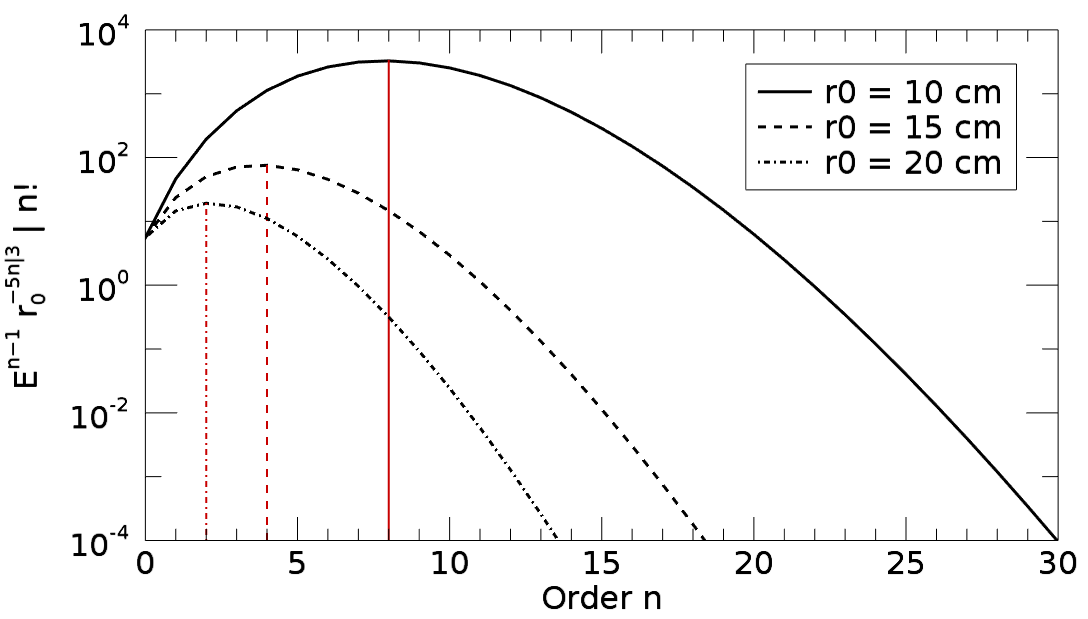}
   \caption{$E^{n-1}r_0^{-5n/3}/n!$ versus $n$, given for $E=0.18$. The different black curves are plotted for different values of the Fried parameter $r_0$. All curves pass by a maximum before decreasing. The colored vertical lines help for visualizing the $N_\mathrm{crit}$ criterion for which the maximum of the curves is reached.}
\label{fig:r0_exp_n}
\end{figure}

The terms of the series are decreasing in magnitude when the following constraint on $n$ is satisfied
\begin{equation}
\label{eq:n_criterion}
\frac{E^{n} r_0^{-5(n+1)/3}}{(n+1)!} < \frac{E^{n-1} r_0^{-5n/3}}{n!} \Longleftrightarrow n > Er_0^{-5/3}-1
\end{equation}
Let's define $N_\mathrm{crit}$ the smaller integer satisfying the inequality. One cannot truncate the sum before $N_\mathrm{crit}$ or would miss terms of higher magnitude. This basic constraint gives a lower bound on the number of terms to be considered in the series for accurate description of the PSF, mathematically we should ensure $N_\mathrm{max}>N_\mathrm{crit}$.The criterion can be visualized on Fig. \ref{fig:r0_exp_n} for different values of $r_0$ and a chosen $E=0.18$ (corresponding to the perfect AO correction up to the cutoff frequency $f_\mathrm{AO}=0.625$ \revision{m$^{-1}$}, see Sec. \ref{sec:AO}). Once again we see that more terms are needed for high phase variance $\sigma^2=B_\phi(0)=Er_0^{-5/3}$. In case of AO correction, \revision{for a given $r_0$}, $E$ depends mainly on the AO cutoff frequency. 

\subsection{Validation on a von-K\'arm\'an turbulence}

\begin{figure}
   \centering
   \includegraphics[width=8cm]{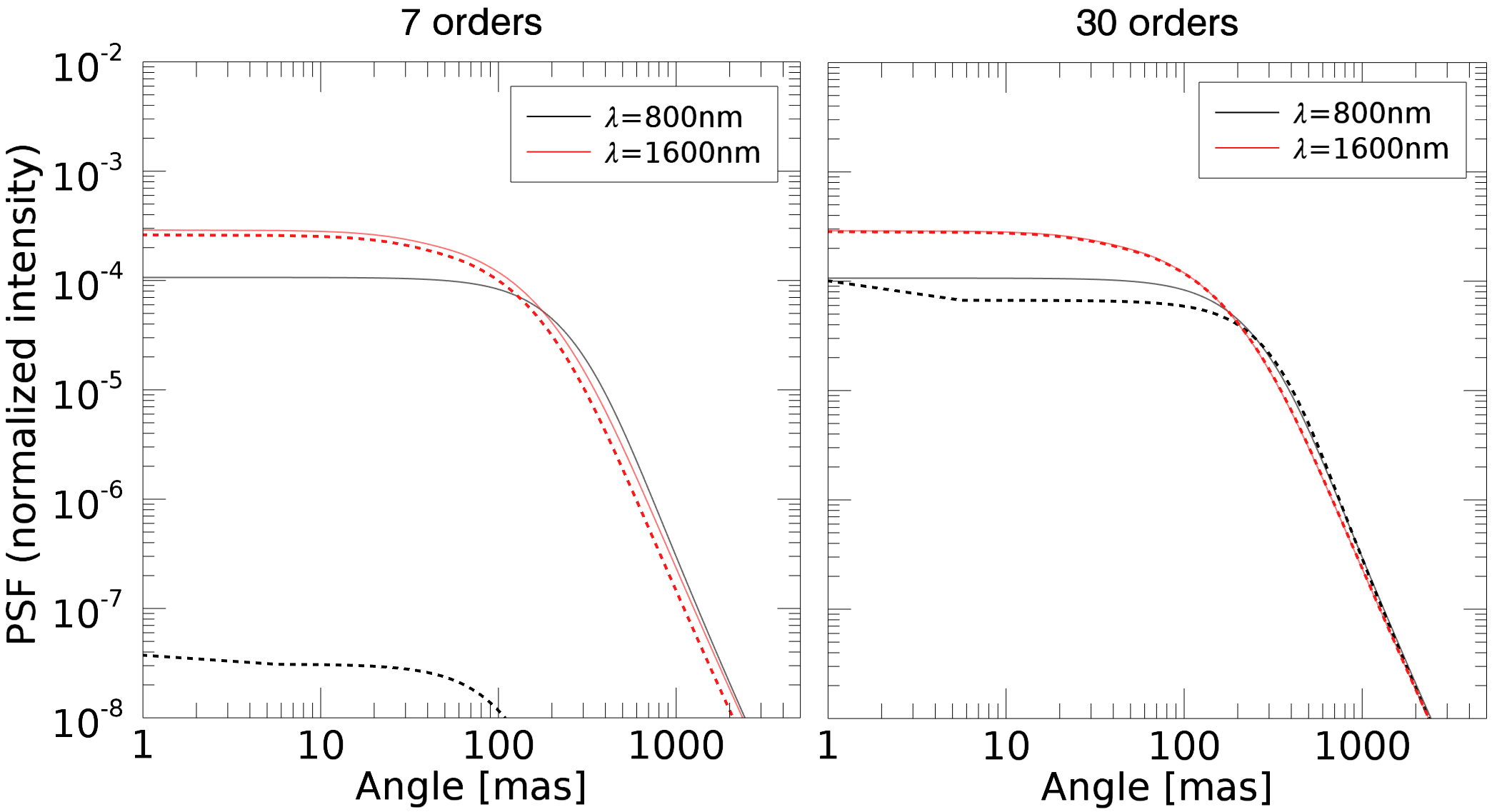}
   \caption{PSF computed from a von-K\'arm\'an spectrum $L_0=8$ m and $r_0=12$ cm at $\lambda=400$ nm. Colored lines show PSF for two wavelengths $\lambda=800$ nm and $\lambda=1600$ nm. Plain: PSF computed by the Roddier OTF method. Dashed: PSF computed by the convolutive orders method. Left panel: 7 orders. Right panel: 30 orders.}
\label{fig:PSF_VK}
\end{figure}

In order to test the validity of the formula, we consider a von-K\'arm\'an spectrum for the turbulent phase, given Eq. (\ref{eq:VK}). On one hand the PSF is generated using the atmospheric OTF as exponential of the phase structure function Eq. (\ref{eq:OTFatmo}). On the other hand, we use our expression giving directly the PSF with the PSD of the phase Eq. (\ref{eq:PSF_as_PSD}). Figure \ref{fig:PSF_VK} shows the PSF for $N_\mathrm{max}=5$ and $N_\mathrm{max}=30$ convolutive orders. For a wavelength of $\lambda=1600$ nm, the convergence requires approximately 7 orders. This is in accordance with our minimal criterion Eq. (\ref{eq:n_criterion}) that gives $N_\mathrm{crit}=4$ at this wavelength. For \revision{a} shorter wavelength ($\lambda=800$ nm) the PSD is higher and the convergence is much slower and requires approximately 30 orders. The criterion gives $N_\mathrm{crit}=22$, it \revision{conforms} the fact that the criterion is \revision{only} a lower bound, but in practice \revision{many} more orders can be required to reach convergence.\\

This test shows the validity of our formula with respect to the \citeauthor{roddier1981v} method on the OTF. This example illustrates the difficulty to approximate the PSF, especially for shorter wavelengths, by the first order of the PSD for a full von-K\'arm\'an spectrum. The method is more suited for low energetic PSD obtained with AO correction, as developed in the following paragraphs.

\subsection{Truncated Kolmogorov \& Adaptive Optics}
\label{sec:AO}

\begin{figure}
   \centering
   \includegraphics[width=8cm]{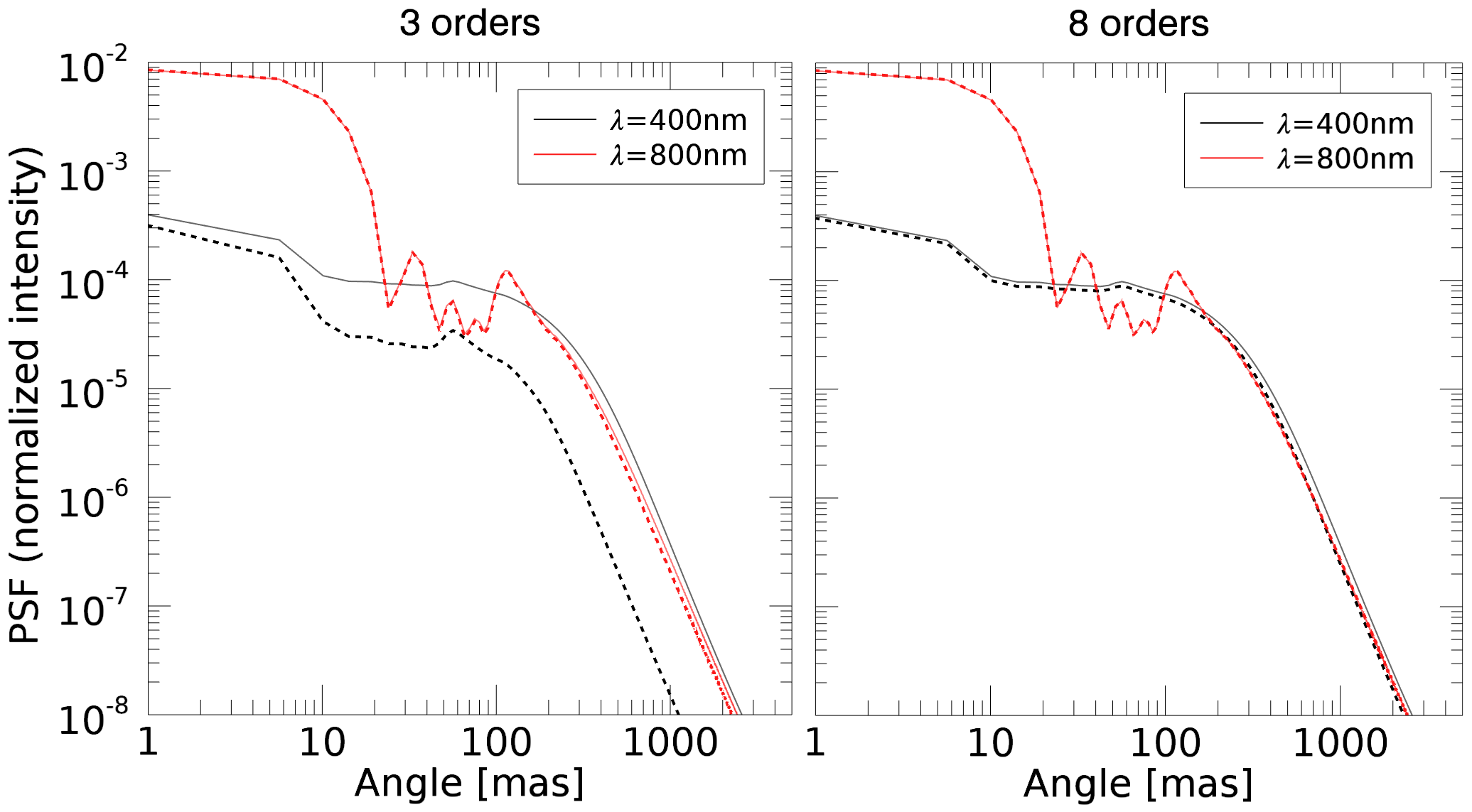}
   \caption{PSF computed for the given AO spectrum. Plain curve: PSF computed by the OTF method. Dashed: PSF computed by the convolutive orders method. Left panel: 3 orders. Right panel: 8 orders.}
\label{fig:PSF_AO}
\end{figure}

\begin{figure}
   \centering
   \includegraphics[width=8cm]{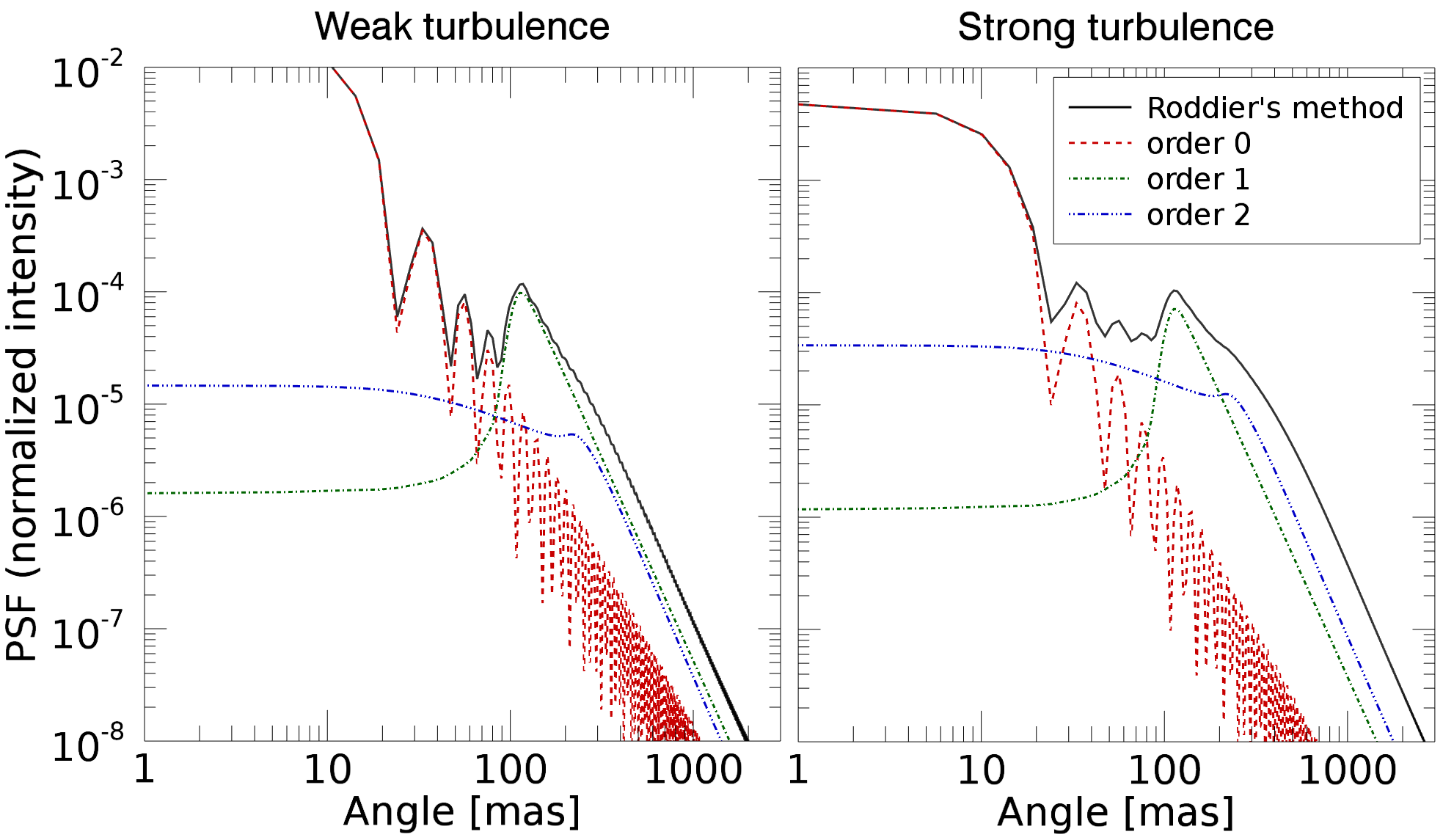}
   \caption{Impact of the $W_\phi$ convolutive orders in the PSF. Each order is also convolved with the pupil PSF, called $h_T$ in the text. Left: $r_0=20$ cm. Right: $r_0=10$ cm. The three orders $n\in\{0,1,2\}$ are respectively plotted in colored dashed lines. The PSF based on the Roddier method, from the OTF in $e^{-D_\phi /2}$, is plotted in plain black.}
\label{fig:Wphi_r0}
\end{figure}

Let's consider a simplified adaptive optics system with a linear number \revision{of actuators $N_{AO}$} in the pupil plane of aperture diameter $D$. All spatial frequencies of the turbulent phase below the AO cutoff frequency $f_{AO}=N_{AO}/(2D)$ are assumed to be perfectly corrected by the AO system. All frequencies above $f_{AO}$ are not seen by the AO system and follow Kolmogorov statistics. The PSD of the residual phase writes
\begin{equation}
W_\phi (f) = 
\left\lbrace
\begin{array}{cl}
0 & \text{ if }f<f_\mathrm{AO}\\
0.023r_0^{-5/3}f^{-11/3} & \text{ if }f\geq f_\mathrm{AO}\\
\end{array}
\right.
\end{equation}
This shape of the AO corrected PSD is plotted in green in Fig. \ref{fig:PSD_models}, for a chosen AO cutoff frequency $f_\mathrm{AO}=0.625$ \revision{m$^{-1}$}, corresponding to $N_{AO}=10$ actuators for a $D=8$ m aperture. In practice, the center of the PSF is dominated by highly energetic low frequency modes, such as residual tip-tilt. Our simple model of a perfect AO high-pass filter doesn't take into account these residues, inducing poor fidelity in the center of the computed PSF. However we can expect a correct description of the halo. Indeed Fig. \ref{fig:PSF_AO} shows good agreements between the two PSF computation methods, using a numerical Fourier transform Eq. (\ref{eq:OTFatmo}) or using the convolutive orders Eq. (\ref{eq:PSF_as_PSD}). Let's notice that the $r_0$ factorization Sec. \ref{sec:sub:r0} is still applicable for numerical computation efficiency with our method on convolutive orders of the PSD. The \revision{atmospheric} PSF is then convolved by $h_T$ to take into account the telescope contribution Eq. (\ref{eq:PSF_tel_atm}). At first order, we retrieve the shape developed in \citet{racine1999speckle} with a coherent core made of the Airy pattern and a turbulent halo depending on the residual PSD. The difference is that our method includes all the convolutive orders of the PSD in the description of the PSF. Higher orders can be of high magnitude when the PSD is of high energy $E$ or for strong turbulence (Fig. \ref{fig:Wphi_r0}). In particular the convolution folds back energy from the PSF halo into the corrected area, reducing contrast in this zone. Magnitude of higher orders can be reduced (and then the contrast in the corrected area increased) by decreasing the integral $E$ of $Y_\phi$, \revision{which} is strongly dependent \revision{on} the cutoff frequency $f_\mathrm{AO}$ and the wavelength $\lambda$ (Fig. \ref{fig:PSF_AO}). Few orders, $N_\mathrm{max}\simeq 3 - 8$ in our example, are required to describe the AO corrected PSF since there is few energy in the \revision{PSD}, to be compared with the $N_\mathrm{max}>30$ orders required to describe the pure von-K\'arm\'an turbulent phase. Our criterion gives respectively $N_\mathrm{crit}=6$ for $\lambda=400$ nm and $N_\mathrm{crit}=1$ for $\lambda=800$ nm. This is in agreement with the PSFs plotted on the figure since the convergence is reached at $N_\mathrm{max}=8$ for $\lambda=400$ nm, and $N_\mathrm{max}=3$ for $\lambda=800$ nm. \\

\subsection{Modeling errors in the AO system}
\label{sec:sub:AOerror}

We previously considered a perfect AO system up to the cutoff frequency, with a phase PSD identically null for $f<f_\mathrm{AO}$. However in practice a residue of phase PSD remains at low frequencies, due to errors of the AO system such as WFS noise, corrective-loop delay, or deformable mirror fitting error. These error terms can be measured or modeled \citep{fusco2006high} for finer description of the AO partial correction. Plugging in these errors in our PSD, we can obtain a better approximation of the PSF, especially in the central peak. Methods directly based on Roddier's OTF formula are already implemented \citep{rigaut1998analytical,jolissaint2010synthetic,correia2017modeling}. Our convolutive formalism then offers \revision{a direct understanding of the link between phase PSD and the PSF} on the results provided by Roddier's OTF based algorithms. \revision{For extreme AO correction we find that the PSF behaves as the phase PSD (first order approximation) whereas the following PSD convolutive orders are required for lower AO correction quality.}\\

We choose here to model the \revision{residual phase PSD due to} AO errors by a Moffat function \citep{moffat1969theoretical}. The shape of the residual PSD writes

\begin{equation}
W_\phi (f) = 
\left\lbrace
\begin{array}{cl}
A(1+f^2/\alpha^2)^{-\beta} + B & \text{ if }f<f_\mathrm{AO}\\
0.023r_0^{-5/3}f^{-11/3} & \text{ if }f>f_\mathrm{AO}\\
\end{array}
\right.
\end{equation}
where $A$, $B$, $\alpha$ and $\beta$ are the Moffat parameters associated to the error budget on the corrected phase. The shape of the PSD for arbitrary values of Moffat parameters is plotted in blue on Fig. \ref{fig:PSD_models}. The resulting PSD is in-between the von-K\'arm\'an uncorrected turbulence, and the perfect AO correction.

\begin{figure}
   \centering
   \includegraphics[width=8cm]{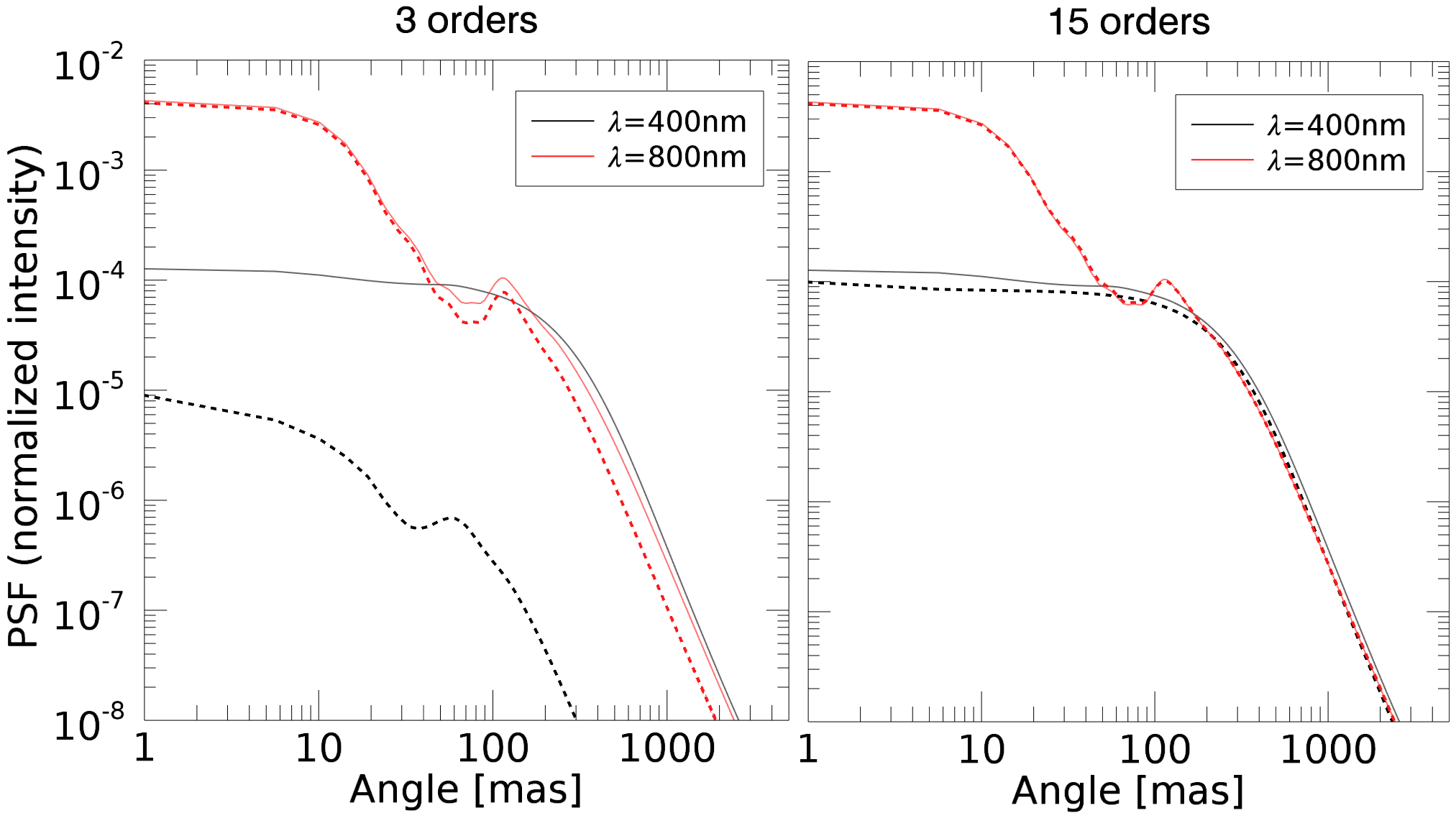}
   \caption{PSF computed for the given AO spectrum. Plain curve: PSF computed by the OTF method. Dashed: PSF computed by the convolutive orders method. Left panel: 3 orders. Right panel: 15 orders.}
\label{fig:PSF_AO_error}
\end{figure}

Figure \ref{fig:PSF_AO_error} shows the convolutive orders associated to the discussed PSD, for $N_\mathrm{max}=3$ and $N_\mathrm{max}=15$. The number of terms required is higher than in the case of a perfect AO system due to a higher energy under the PSD, \revision{but lower than in the case of uncorrected turbulence.}


\section{Conclusions}
\label{sec:conclusion}

For long-exposure observations, the highly time-variable phase aberrations average, and only remains the statistical behavior of the turbulence. We consequently derived a general method to \revision{compute} the long-exposure PSF, given only the PSD of the turbulent phase. When restricting our model to its first order approximation, we find agreement with previous works on the subject \citep{racine1999speckle,bloemhof2001behavior}. We have shown that the higher convolutive orders have more impact on the PSF for stronger turbulence and must be taken into account for low Strehl systems.\\

\revision{A} hypothesis of spatial stationarity of the phase over the pupil is required, that is in agreement with Kolmogorov and von-K\'arm\'an statistics . Our method is also applicable to partially corrected phase with adaptive optics. Finally, we \revision{extracted} the $r_0$ dependency for better understanding of the PSF \revision{shape} with the turbulence strength. It is then possible to get a direct estimation of the $r_0$ given the PSF halo, or reciprocally an estimation of the PSF halo given the $r_0$. We plan to use this model for an accurate description of the halo for AO corrected PSF. The link we make between the PSD and the PSF can be used as a direct measurement of the Fried parameter $r_0$ from a PSF observed on a star. Finally the strong correlation between the AO corrected PSD and the shape of the PSF can provide a tool for AO system diagnostics \citep{jolissaint2002fast}.


\section*{Acknowledgements}

This work was supported by the French Direction G\'en\'erale de l'Armement (DGA) and Aix-Marseille Universit\'e (AMU).\\

This project has received funding from the European Union's Horizon 2020 research and innovation program under grant agreement No 730890. This material reflects only the authors views and the Commission is not liable for any use that may be made of the information contained therein.\\

This study has been partly funded by the French Aerospace Lab (ONERA) in the frame of the VASCO Research Project.



\bibliographystyle{mnras}
\bibliography{references}

\begin{thebibliography}{}
\makeatletter
\relax
\def\mn@urlcharsother{\let\do\@makeother \do\$\do\&\do\#\do\^\do\_\do\%\do\~}
\def\mn@doi{\begingroup\mn@urlcharsother \@ifnextchar [ {\mn@doi@}
  {\mn@doi@[]}}
\def\mn@doi@[#1]#2{\def\@tempa{#1}\ifx\@tempa\@empty \href
  {http://dx.doi.org/#2} {doi:#2}\else \href {http://dx.doi.org/#2} {#1}\fi
  \endgroup}
\def\mn@eprint#1#2{\mn@eprint@#1:#2::\@nil}
\def\mn@eprint@arXiv#1{\href {http://arxiv.org/abs/#1} {{\tt arXiv:#1}}}
\def\mn@eprint@dblp#1{\href {http://dblp.uni-trier.de/rec/bibtex/#1.xml}
  {dblp:#1}}
\def\mn@eprint@#1:#2:#3:#4\@nil{\def\@tempa {#1}\def\@tempb {#2}\def\@tempc
  {#3}\ifx \@tempc \@empty \let \@tempc \@tempb \let \@tempb \@tempa \fi \ifx
  \@tempb \@empty \def\@tempb {arXiv}\fi \@ifundefined
  {mn@eprint@\@tempb}{\@tempb:\@tempc}{\expandafter \expandafter \csname
  mn@eprint@\@tempb\endcsname \expandafter{\@tempc}}}

\bibitem[\protect\citeauthoryear{Bloemhof, Dekany, Troy  \&
  Oppenheimer}{Bloemhof et~al.}{2001}]{bloemhof2001behavior}
Bloemhof E.,  Dekany R.,  Troy M.,   Oppenheimer B.,  2001, The Astrophysical
  Journal Letters, 558, L71

\bibitem[\protect\citeauthoryear{{Conan}}{{Conan}}{1994}]{conan1994etude}
{Conan} J.-M.,  1994, PhD thesis, Universit{\'e} Paris XI Orsay

\bibitem[\protect\citeauthoryear{Correia, Bond, Sauvage, Fusco, Conan  \&
  Wizinowich}{Correia et~al.}{2017}]{correia2017modeling}
Correia C.~M.,  Bond C.~Z.,  Sauvage J.-F.,  Fusco T.,  Conan R.,   Wizinowich
  P.~L.,  2017, JOSA A, 34, 1877

\bibitem[\protect\citeauthoryear{Fried}{Fried}{1966}]{fried1966optical}
Fried D.~L.,  1966, JOSA, 56, 1372

\bibitem[\protect\citeauthoryear{Fusco et~al.,}{Fusco
  et~al.}{2006}]{fusco2006high}
Fusco T.,  et~al., 2006, Optics Express, 14, 7515

\bibitem[\protect\citeauthoryear{Goodman}{Goodman}{1968}]{goodman1968fourier}
Goodman J.~W.,  1968, Appendix B Sec. B, 3

\bibitem[\protect\citeauthoryear{{Jolissaint}}{{Jolissaint}}{2010}]{jolissaint2010synthetic}
{Jolissaint} L.,  2010, Journal of the European Optical Society -- Rapid
  Publications, 5, 10055

\bibitem[\protect\citeauthoryear{Jolissaint \& Veran}{Jolissaint \&
  Veran}{2002}]{jolissaint2002fast}
Jolissaint L.,  Veran J.-P.,  2002, ESO Conference and Workshop Proceedings.
 Vol. 58

\bibitem[\protect\citeauthoryear{Moffat}{Moffat}{1969}]{moffat1969theoretical}
Moffat A.,  1969, Astronomy and Astrophysics, 3, 455

\bibitem[\protect\citeauthoryear{Perrin, Sivaramakrishnan, Makidon, Oppenheimer
   \& Graham}{Perrin et~al.}{2003}]{perrin2003structure}
Perrin M.~D.,  Sivaramakrishnan A.,  Makidon R.~B.,  Oppenheimer B.~R.,
  Graham J.~R.,  2003, The Astrophysical Journal, 596, 702

\bibitem[\protect\citeauthoryear{Racine, Walker, Nadeau, Doyon  \&
  Marois}{Racine et~al.}{1999}]{racine1999speckle}
Racine R.,  Walker G.~A.,  Nadeau D.,  Doyon R.,   Marois C.,  1999,
  Publications of the Astronomical Society of the Pacific, 111, 587

\bibitem[\protect\citeauthoryear{Rigaut, V{\'e}ran  \& Lai}{Rigaut
  et~al.}{1998}]{rigaut1998analytical}
Rigaut F.~J.,  V{\'e}ran J.-P.,   Lai O.,  1998, Proc. SPIE.
 Vol. 3353, pp 1038--1048

\bibitem[\protect\citeauthoryear{Roddier}{Roddier}{1981}]{roddier1981v}
Roddier F.,  1981, in , Vol.~19, Progress in optics.
Elsevier, pp 281--376

\bibitem[\protect\citeauthoryear{Roddier}{Roddier}{1999}]{roddier1999adaptive}
Roddier F.,  1999, Adaptive optics in astronomy.
Cambridge university press

\bibitem[\protect\citeauthoryear{Sivaramakrishnan, Lloyd, Hodge  \&
  Macintosh}{Sivaramakrishnan et~al.}{2002}]{sivaramakrishnan2002speckle}
Sivaramakrishnan A.,  Lloyd J.~P.,  Hodge P.~E.,   Macintosh B.~A.,  2002, The
  Astrophysical Journal Letters, 581, L59

\makeatother
\end{thebibliography}



\begin{appendix}

\section{On the convolutive approximation}
\label{appendix:approx}

\begin{figure}
   \centering
   \includegraphics[width=7cm]{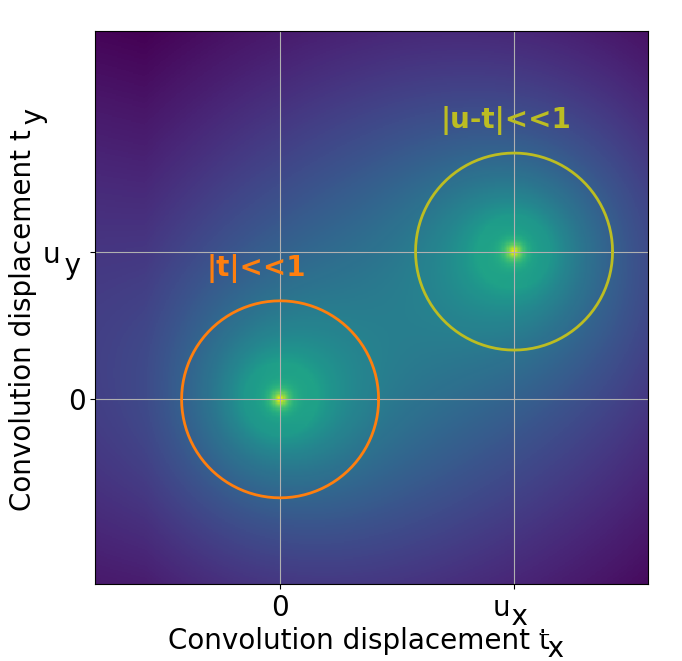}
   \caption{Visualization of the three different domains for a given convolutive shift $\vec{u}$. Colormap show the product $Y_\phi(\vec{t})Y_\phi(\vec{u}-\vec{t})$. Intensity plot is in logarithmic scale.}
\label{fig:convol_proof}
\end{figure}

\revision{

Let's consider a PSD shape verifying the two following hypothesis
\begin{enumerate}
\item The majority of the PSD energy is in the center\\ ($|\vec{u}|\ll 1$) rather than in the halo ($|\vec{u}|\gg 1$).
\item The halo is slowly varying.
\end{enumerate}
These hypothesis are verified in particular for the $-11/3$ power law of the Kolmogorov spectrum.

Let's now consider $\vec{u}$ a coordinate in the halo. The PSD auto-convolution then writes
\begin{equation}
Y_\phi \star Y_\phi (\vec{u}) = \iint_{\mathbb{R}^2}Y_\phi(\vec{t})Y_\phi(\vec{u}-\vec{t}) d\vec{t}
\end{equation}
The $\mathbb{R}^2$ domain is then separated into three subdomains
\begin{itemize}
\item $|\vec{t}|\ll 1$ where $Y_\phi(\vec{t})$ is large due to the first hypothesis, and $Y_\phi(\vec{u}-\vec{t})$ is slowly varying due to the second one (when choosing $\vec{u}\gg 1$)
\item $|\vec{u}-\vec{t}|\ll 1$ where the opposite behavior comes out
\item The remaining domain of $\mathbb{R}^2$, called $\mathcal{A}$, where both $Y_\phi(\vec{t})$ and $Y_\phi(\vec{u}-\vec{t})$ are small due to  the first hypothesis
\end{itemize}
Figure \ref{fig:convol_proof} shows an example of these domains for a given $\vec{u}$ and a PSD shape as described in Sec. \ref{sec:sub:AOerror}. The convolutive integral of $Y_\phi$ is separated on the three domains described above. The integral on the first domain is
\begin{equation}
\iint_{|\vec{t}| \ll 1}Y_\phi(\vec{t})Y_\phi(\vec{u}-\vec{t}) d\vec{t} \simeq Y_\phi(\vec{u}) \iint_{|\vec{t}| \ll 1}Y_\phi(\vec{t}) d\vec{t}
\end{equation}
The approximation we just made rely on the fact that $|\vec{u}|\gg 1$ is in the halo and $|\vec{t}| \ll 1$ so $\vec{u}-\vec{t}$ is still in the halo. The slow variations in the halo (second hypothesis) allows to approximate $Y_\phi(\vec{u}-\vec{t})\simeq Y_\phi(\vec{u})$. We make the same reasoning on the second domain, with a coordinate change $\vec{x}=\vec{u}-\vec{t}$, so we find
\begin{equation}
\iint_{|\vec{u}-\vec{t}| \ll 1}Y_\phi(\vec{t})Y_\phi(\vec{u}-\vec{t}) d\vec{t} \simeq Y_\phi(\vec{u}) \iint_{|\vec{x}| \ll 1}Y_\phi(\vec{x}) d\vec{x}
\end{equation}
The two first integrals are then identical. Regarding the third domain, due to the first hypothesis, it is a second order terms as the convolution of two small terms. We simply write it
\begin{equation}
\iint_{\mathcal{A}}Y_\phi(\vec{t})Y_\phi(\vec{u}-\vec{t}) d\vec{t}  = \eta(\vec{u}) E
\end{equation}
where $\eta(\vec{u})$ is a small valued function, and $E$ denotes the integral of $Y_\phi$ on $\mathbb{R}^2$. Summing the partial convolutions on the three domains, one gets the full convolution
\begin{equation}
Y_\phi \star Y_\phi (\vec{u})  \simeq 2 Y_\phi(\vec{u}) \left[ \iint_{|\vec{t}| \ll 1}Y_\phi(\vec{t}) d\vec{t} \right] + \eta(\vec{u}) E
\end{equation}
Finally, using the first hypothesis, the integral above contains the majority of the energy. Then we write
\begin{equation}
\iint_{|\vec{t}| \ll 1}Y_\phi(\vec{t}) d\vec{t}  = \kappa E
\end{equation}
where $\kappa\in ]0,1]$ represents the ratio of energy in the area $|\vec{t}| \ll 1$. It depends on the shape of $Y_\phi$, in practice we find $\kappa\simeq 0.85$ for our AO truncated Kolmogorov. Using $\kappa$, the convolution rewrites
\begin{equation}
Y_\phi \star Y_\phi (\vec{u})  \simeq 2\kappa Y_\phi(\vec{u}) E + \eta(\vec{u}) E
\end{equation}
If we neglect the $\eta(\vec{u}) E$ term, one can easily iterate by recurrence over the convolutive orders to find
\begin{equation}
\{ \star^n~ Y_\phi \} (\vec{u})  \simeq (2\kappa E)^{n-1} Y_\phi(\vec{u})
\end{equation}
Errors in this equation come from the different approximations we made and the modification of shape for $\{ \star^n~ Y_\phi \}$, especially in the center ($\vec{u}\ll 1$). One should remind that this expression is only true in order of magnitude since the approximations propagate through the convolutive orders.

}

\end{appendix}


\bsp	
\label{lastpage}
\end{document}